\documentclass[aps,showpacs,twocolumn]{revtex4}
\usepackage{amsfonts}
\usepackage{amsmath}
\usepackage{amssymb}
\usepackage{graphicx}
\usepackage{bm}

\input{tcilatex}

\begin{document}

\title{Leidenforst gas ratchets driven by thermal creep}
\author{Alois W\"{u}rger}
\affiliation{Laboratoire Ondes et Mati\`{e}re d'Aquitaine, Universit\'{e} de Bordeaux \&
CNRS, 351 cours de la Lib\'{e}ration, 33405 Talence, France}

\begin{abstract}
We show that thermal creep is at the origin of the recently discovered
Leidenfrost ratchet, where liquid droplets float on a vapor layer along a
heated saw-tooth surface and accelerate to velocities of up to 40 cm/s. As
the active element, the asymmetric temperature profile at each ratchet
summit rectifies the vapor flow in the boundary layer. This mechanism works
at low Reynolds number and provides a novel tool for controlling gas flow at
nanostructured surfaces.

\end{abstract}

\maketitle

Liquid spilled on a hot surface rapidly evaporates. At the Leidenfrost
temperature well above the boiling point, however, one observes long-lived
droplets that levitate due to the excess pressure resulting from the
permanent feed of vapor at the bottom. Their contact-free suspension makes
such Leidenfrost droplets very mobile. Linke et al. observed that, when
placed on a millimeter-sized brass ratchet, the droplets rapidly accelerate
to\ a speed of about 10 cm/s \cite{Lin06}. Very recently, Lagubeau et al.
found that the same effect occurs for a piece of solid dry ice, and thus is
not related to properties of the liquid phase \cite{Lag11}. Even more
surprisingly, Ok et al. reported that reducing the ratchet profile to 200
nm, has little effect on the droplet velocity \cite{Ok11}.

Unlike other self-propulsion mechanisms based on chemical or thermal
gradients \cite{Bro89,Caz90,deG03,Cha92,Ich00}, this motion is not directed
along an applied field but rather arises from the asymmetric surface
structure of the solid support. This sawtooth profile acts as a rectifier
transferring momentum on the interstitial vapor; the resulting gas flow
advects the floating Leidenfrost droplet. The experimental findings \cite%
{Lin06,Lag11,Ok11} suggest that there is a common principle that works for
both liquids and solids, and independently of the height of the ratchet
profile. Several ideas have been put forward, relying on non-uniform Laplace
pressure and Marangoni forces in the droplet, surface vibrations, or
rectification of the radial vapor flow through the non-linear term of the
Navier-Stokes equation \cite{Lin06,Lag11,Ok11,Que06}; yet none of them
explains all of the mentioned experiments.\ In particular, the submicron
ratchets of Ok et al. \cite{Ok11} exclude non-linear hydrodynamics as the
dominant mechanism, as illustrated in the left panel of Fig. 1: For small
profile $D$, the gas velocity and the effective Reynolds number $\func{Re}$
in the ratchet layer are much smaller than at midheight where $\func{Re}\sim
1$ \cite{Lag11}; thus rectification is expected to disappear for $D\ll h_{0}$%
, whereas the data of Ok et al. rather show the opposite behavior.

\begin{figure}
\includegraphics[width=\columnwidth]{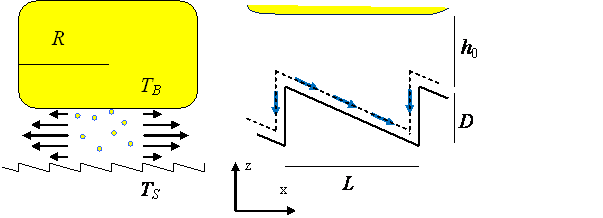}
\caption{Leidenfrost droplet on a
ratchet. The left panel shows the gas flow due to the evaporation at the
bottom of a droplet of radius $R$; the temperature of the solid $T_{S}$ is
significantly above the droplet's boiling temperature $T_{B}$. The arrows in
the right panel indicate the thermal creep flow along the ratchet of period $%
L$ and height $D$. The dashed line indicates the boundary layer; its
thickness $\ell $ corresponds to the molecular mean free path. In recent
experiments the ratchet height $D$ varies from 200 nm to about 1 mm; for
small $D$ one has $h_{0}=10...100\protect\mu $m. }
\end{figure}

In his 1879 attempt to explain
Crookes' radiometer experiment and building on Reynolds' theory for thermal
transpiration \cite{Rey79}, Maxwell showed the existence of thermal creep
velocity along a solid-gas interface \cite{Max79}, 
\begin{equation}
v_{C}=\frac{3}{4}\nu \frac{\nabla _{\parallel }T}{T},  \label{2}
\end{equation}%
where $\nu $ is the kinematic viscosity of the vapor and $\nabla _{\parallel
}T$ the parallel component of the temperature gradient. Kinetic theory
relates this gas flow to the non-uniform density and velocity distribution:
Molecules coming from the cold side and hitting the surface at a given
point, are more frequent but carry lower momentum than those from the hot
side, thus resulting in an off-diagonal component of the surface stress and
the boundary velocity $v_{C}$ \cite{Max79}. Thermal creep drives aerosol
thermophoresis \cite{Der65}, repels air-suspended particles from a hot
surface \cite{Tal80}, and operates in small-scale gas flow devices such as
thermally actuated microcantilevers and Knudsen pumps \cite%
{Var99,Pas03,Col05,Got05,Nam05,Gup08}.

The present Letter points out the role of thermal creep for self-propelling
Leidenfrost droplets and, in particular, analyzes the flow around a ratchet
summit. The essential argument is illustrated in the right panel of Fig. 1.
In the cleft below the droplet, there is a strong temperature gradient of
several tens of Kelvin per micron. Because of the asymmetric profile, the
horizontal component of the creep velocity has a finite mean value; the
resulting gas flow along the ratchet surface drags the droplet toward the
right. Note that this argument does not rely on the existence of the outward
gas flow illustrated in the left panel of Fig.\ 1.

Our detailed analysis relies on Stokes hydrodynamics. In anology to thermal
transport in colloidal dispersions \cite{Pia08,Wue10}, the droplet velocity
is derived from the overall force balance on a closed surface. In the
absence of external forces in horizontal direction one has 
\begin{equation}
\oint \sigma _{xn}dS=0,  \label{6}
\end{equation}%
where $\sigma _{xn}$ is the stress pulling in $x$-direction on the area
element $dS$ with normal $n$. The stress tensor $\sigma _{ij}=\sigma
_{ij}^{\prime }-P\delta _{ij}$ comprises a viscous part $\sigma
_{ij}^{\prime }=\eta (\partial _{i}v_{j}+\partial _{j}v_{i})$ and the excess
pressure $P$. A non-uniform flow velocity $v$ in the cleft of width $h$ \
creates a stress of the order $\eta v/h$, which by far exceeds the viscous
drag at the remaining part of the droplet surface $\sim \eta v/R$. Thus the
surface integral may be limited to the part between droplet and support; it
closely follows the ratchet profile beyond the boundary layer, as indicated
by the dashed line in Fig. 1.

The velocity profile in the cleft comprises two contributions of different
origin. The first one, due to evaporation at the bottom of the droplet, is
the outside gas flow in the left panel of Fig.\ 1; in the framework of
Stokes hydrodynamics it does not contribute to the stress integral. Thus in
the following, we consider the second velocity term, which arises from the
thermal creep along the ratchet profile, as indicated by the arrows in the
right panel.

The rectification mechanism is most obvious when comparing the viscous
stress at the two slopes of the ratchet. The normal on the vertical part
points in $x$ direction; the corresponding diagonal element $\sigma
_{xx}^{\prime }=2\eta dv_{x}/dx$ vanishes since $v_{x}$ and its derivative
are zero. On the opposite side of slope $m=D/L$, the stress $\sigma
_{xn}^{\prime }$ is finite. The hydrostatic pressure varies little along the
profile and will be discarded; then Eq. (\ref{6}) reduces to the condition%
\begin{equation}
\left\langle \sigma _{xz}^{\prime }\right\rangle =\frac{1}{L}%
\int_{0}^{L}\sigma _{xz}^{\prime }dx=0  \label{8}
\end{equation}%
on the viscous drag on the slope of the ratchet tooth. If the droplet is
immobile, the shear stress reads as $\sigma _{xn}^{\prime }\approx -\eta
v_{C}/h$, where $h$ is the width of the cleft. In order to satisfy (\ref{8})
the droplet moves at a velocity $u$, leading to $\sigma _{xz}^{\prime }=\eta
\lbrack u-v_{C}(x)]/h$. Inserting $\sigma _{xz}^{\prime }$ in (\ref{8}) one
readily obtains the expression for the drift velocity, 
\begin{equation}
u=\frac{\left\langle v_{C}/h\right\rangle }{\left\langle 1/h\right\rangle }.
\label{12}
\end{equation}

The temperature profile is determined by the boundary conditions at the
solid-gas interface, imposing continuous temperature and heat flow through
the interface.\ Because of the important difference in thermal conductivity
of the brass support and the vapor layer, $\kappa _{V}/\kappa _{S}\sim
10^{-4}$, the temperature profile is strongly distorted, and the gradient is
much larger in the vapor phase.

\begin{figure}
\includegraphics[width=\columnwidth]{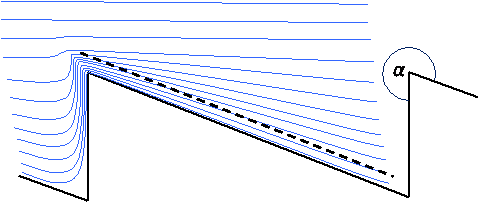}
\caption{Temperature profile close to a solid-vapor interface. Due to the small
conductivity ratio $\protect\kappa _{V}/\protect\kappa _{S}$, the isotherms
(solid lines) are strongly distorted. At a distance of one molecular mean
free path from the hot surface (dashed line), the temperature gradient in
the vapor has a significant component parallel to the surface, $\protect%
\nabla T_{\parallel }$, which is largest close to the upper corner. }
\end{figure}

For a first estimate we calculate\textit{\ }$\nabla T_{\parallel }$ far from
the corners, in the middle part of a ratchet tooth. In the limit $\kappa
_{V}/\kappa _{S}\rightarrow 0$ the brass surface is at constant temperature $%
T_{S}$, and the profile in the cleft is given by $T(x,z)=T_{B}-(z/h)\Delta T$%
, where $\Delta T=T_{S}-T_{B}$ and $h=h_{0}+xD/L$. The resulting temperature
gradient is perpendicular on the solid-vapor and droplet-vapor interfaces.
The velocity distribution of the molecules hitting the brass surface is
given by the temperature profile evaluated at one mean-free path from the
brass surface, $z=\ell -h$. At this finite distance, the gradient has a
component parallel to the surface $\nabla T_{\parallel }\sim \Delta T\ell
/h^{2}$. Replacing $h$ with $h_{0}$ and discarding numerical factors gives a
rough estimate for the drift velocity, 
\begin{equation}
u\sim \nu \ell /h_{0}^{2}.  \label{14}
\end{equation}%
With the mean-free path $\ell =130$ nm and the kinematic viscosity $\nu =60$
mm$^{2}$/s \ of vapor at 300$%
{{}^\circ}%
$ C, and $h_{0}\sim 10\mu $m \cite{Bia03},\ one finds $u\sim 10$ cm/s, which
is in qualitative agreement with experiment \cite{Lin06,Lag11,Ok11}.

For the sake of a more quantitative description we refine the vapor
temperature profile close to the upper corner of the ratchet, which turns
out to dominate the creep flow. In analogy to the electrostatic potential of
a charged polygon, a simple conformal transformation provides the expression 
$T(r,\varphi )=T_{S}-\Delta T(r/h_{0})^{\pi /\alpha }\sin (\pi \varphi
/\alpha )$ \cite{Dri02}, where $r,\varphi $ are polar coordinates with
respect to the corner.\ The angle $\alpha $ is related to the aspect ratio $%
m=D/L=-\cot \alpha $; for the ratchets of Ref. \cite{Ok11} one finds the
exponent $\pi /\alpha \approx 0.63$. The resulting parallel component of the
gradient along the dashed line close to point $A$\ reads \cite{suppmat}%
\begin{equation}
\nabla T_{\parallel }=\hat{\xi}\Delta T\frac{\ell h_{0}^{-\pi /\alpha }}{%
r^{2-\pi /\alpha }},
\end{equation}%
where $\hat{\xi}=-\frac{\pi ^{2}}{\alpha ^{2}}\cos \frac{\alpha }{2}$. Its
essential feature is the weak singularity at the ratchet summit, very
similar to the electric field close to a charged cusp. The molecular
mean-free path $\ell $ provides a physical cut-off for the divergency.

\begin{figure}
\includegraphics[width=\columnwidth]{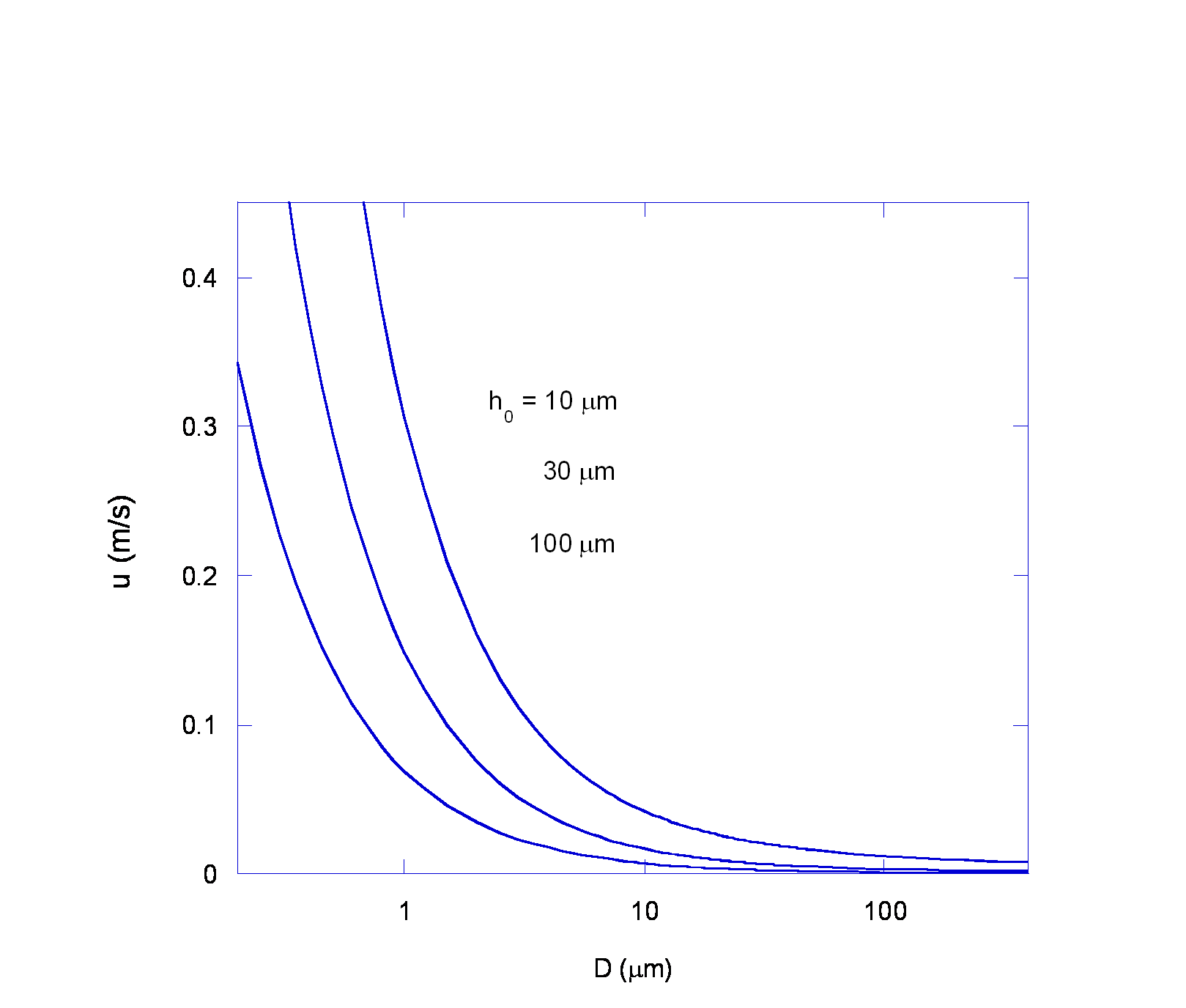}
\caption{Drift velocity $u$ as a
function of the ratchet parameter $D$ for $h_{0}=10,30,100\protect\mu $m.
The curves are calculated from Eq. (\protect\ref{16}) with $\protect\nu =60$%
mm$^{2}$/s, $\ell =130$ nm, $\protect\pi /\protect\alpha =0.63$, $L/D=4$,
and $\Delta T/T=1/2$. }
\end{figure}

Now the drift velocity is evaluated in terms of Eq. (\ref{12}), resulting in 
\cite{suppmat} 
\begin{equation}
u=\xi \frac{\Delta T}{T_{S}}\frac{\nu }{h_{0}}\left( \frac{\ell }{h_{0}}%
\right) ^{\pi /\alpha }\frac{D/L}{\ln (1+D/h_{0})}  \label{16}
\end{equation}%
with the numerical prefactor $\xi \approx 0.6$ \cite{suppmat}. This
expression confirms the estimate (\ref{14}) yet shows additional
dependencies on the ratchet parameters. Fig. 3 reveals a striking variation
of $u$ with $D$: the smaller the ratchet profile, the larger the droplet
velocity. This at first sight counterintuitive result is confirmed by the
experiment of Ok et al. \cite{Ok11}: Their data at intermediate temperatures
are well fitted by a logarithmic variation, similar to (\ref{16}). Though
this comparison does not account for the implicit dependence of $h_{0}$ on $%
D $, one may safely conclude on a qualitative agreement of (\ref{16}) with
the data. Note that for a ratchet driven by non-linear hydrodynamics, one
expects the opposite behavior, i.e., a smaller velocity for small $D$.
Indeed, from the left panel of Fig. 1 it is clear that for $D\ll h_{0}$, the
gas velocity in the ratchet layer is small and the non-linear term $(\mathbf{%
v}\cdot \mathbf{\nabla })\mathbf{v}$ of the Navier-Stokes equation
insignificant.

So far we have considered liquid droplets on a ratchet. The same mechanism
holds for a piece of dry ice (solid CO$_{2}$) above its sublimation
temperature floating above a hot metal surface; when graving a saw tooth
profile at its lower face, Lagubeau et al. observed motion similar to the
droplets discussed so far. Since the creep velocity occurs at the bottom of
the dry ice, the mean velocity of the vapor in the cleft is zero, as
illustrated in Fig.\ 4.

\begin{figure}
\includegraphics[width=\columnwidth]{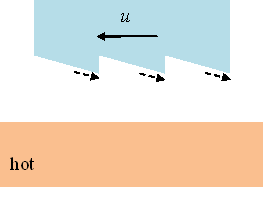}
\caption{Thermal
creep below a piece of dry ice (solid CO$_{2}$) floating above a hot metal
surface due to sublimation \protect\cite{Lag11}. Due to the ratchet profile
printed at its lower side, there is a parallel temperature gradient $\protect%
\nabla T_{\parallel }$ as indicated by dashed arrows; the rectified thermal
creep flow propels the disk to the left. The mean velocity of the vapor in
the cleft is zero. }
\end{figure}

Lagubeau et al. measured the force $F$ which required to immobilize a
droplet floating on a ratchet \cite{Lag11}. In the range $R=1...7$ mm, they
found values from 3 to 30 microNewton, and a power law $F\propto R^{\beta }$%
, with an exponent $\beta \approx 1.5$. In the present work, this force is
given by the integral of the shear stress over the contact area, $F=\pi
R^{2}\eta \left\langle v_{C}/h\right\rangle $. With the above expression for
the thermal creep velocity one finds 
\begin{equation}
F=\pi \xi \frac{\eta R^{2}D}{L^{2}}\nu \frac{\Delta T}{T}\frac{\ell ^{\pi
/\alpha }}{h_{0}^{1+\pi /\alpha }}.
\end{equation}%
With the relation $h_{0}\propto \sqrt{R}$ \cite{Lag11}, the force varies
with the droplet size as $\beta =\frac{3}{2}-\frac{\pi }{2\alpha }\approx
1.2 $; within the experimental uncertainities, this compares favorably with
the measured value.

The thermal-creep mechanism \ described here is not limited to the motion of
Leidenfrost droplets. As a straightforward application we discuss the gas
pump shown in Fig. 5, which consists of two nanostructured plates at
temperatures differing by $\Delta T$. Both solid interfaces show thermal
creep and thus impose a unifom gas flow across the cleft.\ For a
sufficiently small ratchet profile, $D<h_{0}$, Eq. (\ref{16}) simplifies to 
\begin{equation}
u_{0}=\xi \frac{\nu }{L}\frac{\Delta T}{T}\left( \frac{\ell }{h_{0}}\right)
^{\pi /\alpha }.  \label{18}
\end{equation}%
This velocity may attain several meters per second. It turns out instructive
to compare this ratchet with a Knudsen pump; in the present case, the
thermal gradient is perpendicular to the gas flow, whereas both are parallel
in the latter. Moreover, a Knudsen pump requires the system size to be
comparable to or smaller than the mean-free path, and thus is restricted to
very dilute gases. Although the ratchet mechanism does depend on the ratio $%
\ell /h_{0}$, it works for films that are hundred times thicker than the
mean free path.

\begin{figure}
\includegraphics[width=\columnwidth]{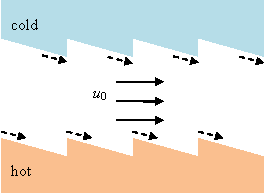}
\caption{Gas pump driven
by thermal creep. The thermal gradient across the channel is given by their
temperature difference $\Delta T$ and spacing $h_{0}$. At the ratchet
corners, there is a parallel component $\protect\nabla T_{\parallel }$, as
indicated by dashed arrows. Thermal creep gives rise to a uniform gas flow
at velocity $u_{0}$, as given in Eq. (\protect\ref{12}). Note the opposite
orientation of the saw teeth on the cold and hot side.}
\end{figure}

\end{document}